\documentstyle[12pt]{article}

\begin{document}

\topmargin 0pt
\oddsidemargin 5mm

\setcounter{page}{0}
\begin{titlepage}
\vspace{2cm}
\begin{center}
{{\large\bf
 STRUCTURE CONSTANTS IN THE $N=1$ SUPER-LIOUVILLE FIELD
THEORY
}}\\

\vspace{1cm}

{\large R.H.Poghossian}\\
\vspace{1cm}

{\em Yerevan Physics Institute,  Republic of Armenia}\\
{(Alikhanian Brothers St. 2,  Yerevan 375036,  Armenia)}\\
{e-mail: poghossian@vx1.yerphi.am}
\end{center}

\vspace{5mm}

\centerline{{\bf{Abstract}}}

The symmetry algebra of $N=1$ Super-Liouville field theory in two
dimensions is the infinite dimensional $N=1$ superconformal algebra,
which allows one to prove, that correlation functions, containing
degenerated fields obey some partial linear differential equations.
In the special case of four point function, including a primary
field degenerated at the first level, this differential equations
can be solved via hypergeometric functions. Taking into account
mutual locality properties of fields and investigating s- and t-
channel singularities we obtain some functional relations for three-
point correlation functions. Solving this functional equations we
obtain three-point functions in both Neveu-Schwarz and Ramond
sectors.

\vfill
\centerline{\large Yerevan Physics Institute}
\centerline{\large Yerevan 1996}

\vfill
\end{titlepage}

\newpage

\section{Introduction}
\vspace{3mm}

Attempts to achieve deeper understanding of two dimensional
quantum (super-) gravity \cite{P1}-\cite{P7} are mainly
motivated by the fact, that these theories appear to be one of the
most important building blocks of noncritical (super-) strings
\cite{P8,P9}.

The (super-) Liouville field theory, which is the effective
theory of (super-) gravity in two dimensions, has an infinite
dimensional (super-) conformal invariance inherited from the
initial general (super-) covariance. Therefore, it might be
possible to use powerful methods of two dimensional Conformal
Field Theory (CFT)
 to investigate these theories. The main
difficulty in this direction is due to the facts, that the
physical Hilbert space of (super-) Liouville field theory contains
infinite continuum set of primary states and the local
field$\leftrightarrow$state correspondence, which is usual in CFT,
is problematic. As
has been shown in \cite{P12} the correlation functions in
Liouville field theory
can be calculated using a technique, quite similar to
the Coulomb gas representation in ordinary "minimal" models of CFT
\cite{P13}, provided some on mass-shell type conditions are
satisfied.
In \cite{P14,P15}, \cite{P16} the three point functions of
exponential fields in Liouville field theory have been calculated
first in above mentioned on mass-shell case, after which a general
expression has been conjectured. But there is another method of
calculating three point functions in CFT too, which 
has been used successfully for several models of CFT in \cite{P17, P18}.
The method is the following. As it is well known
\cite{P10}, correlation functions, containing at least one degenerated
field, obey some linear differential equations. In many
interesting cases, when degeneration takes place at low levels,
four-point correlation functions can be obtained directly solving
these differential equations,  thus avoiding Coulomb gas type
representations. The latter is essential, because neutrality
condition, necessary for having Coulomb gas representation, touches all
the fields entering correlation function  contrary to the
degeneracy condition, which is related to a separate field only.
This is the reason why investigating four point functions with
the help of Coulomb gas representation one obtains relations only
for on mass-shell  three point functions, while investigation of four
point functions, containing a single
degenerated field, leads to some nontrivial functional relations
for unconstrained three point functions. In the cases of minimal
models these relations are reduced to   recurrent equations,
solving which one obtains three point functions
\cite{P17}, \cite{P18}. In the case of Liouville field theory,
solving above mentioned functional relations, all the results,
conjectured in \cite{P15, P16} are reobtained in \cite{P19}. In
this work, the same method is used in the case of $N=1$ Supper-
 Liouville Field theory (SLFT).
  The further part of this paper is organized  as follows.
  In sec.2 we present a brief review on $N=1$ SLFT. In sec.3
  it is shown that four point correlation functions, including a
  Ramond field degenerated at second level, obey some linear
  differential equation. Solving these equations and taking into
  account locality condition we express four point correlation
  functions via hypergeometric functions.
  In sec.4 investigating s- and t-channel singularities of four
  point correlation functions obtained in sec.3, some functional
  relations for structure constants are derived.
  Solving these functional relations
  all three point functions of
 exponential fields and reflection amplitudes are calculated in
 both Neveu-Schwarz and Ramond sectors.

The results of this work have been previously reported in
\cite{P20}

\section{\bf $N=1$ Super-Liouville, as a Two Dimensional Superconformal
Field Theory}
\vspace{3mm}

The Super-Liouville field theory is a supersymmetric generalization
of the bosonic Liouville theory, which is known to be the theory of
matter induced gravity in two dimensions. Similarly SLFT describes 2d
supergravity, induced by supersymmetric matter. To obtain Super-Liouville
action, one can simply "supersymmetrize" the bosonic Liouville action.
The answer reads
\begin{equation}
\label{A1}
S_{SL}=\frac{1}{4\pi }\int \hat{E}\left[ \frac{1}{2}D_\alpha
\Phi_{SL}
D^\alpha \Phi_{SL} -Q\hat{R }\Phi_{SL} -4\pi i\mu e^{b\Phi_{SL}}
\right],
\end{equation}
where $\Phi_{SL}$ is Liouville superfield, $D^{\alpha },D_{\alpha }$
are superderivatives, $\hat{Y}$
and $\hat{E}$ are supercurvature and superdensity of the background super\-manifold  (see \cite{P5}). The condition, that cosmological term $\mu
\exp b\Phi_{SL}$ has correct
dimension $(1/2,1/2)$ leads to the following relation between "background
charge" $Q$ and the coupling constant $b$
\begin{equation}
\label{A2}
Q=b+\frac{1}{b}.
\end{equation}

One of the most important properties of the action (2.1) is its 
superconformal invariance. To describe superconformal symmetry and its
consequences more explicitly, let us consider SLFT on the superplane,
with coordinates $(Z,\bar{Z})=(z,\bar{z},\theta,\bar{\theta })$, where
$z,\bar{z}$ are complex coordinates on plain,
and $\theta, \bar{\theta}$ are corresponding  Grassmanian coordinates
(for topologically
nontrivial supermanifolds such choice of "flat" coordinates can be achieved
for every coordinate patch separately, using superdiffeomorfism and super-
Weyl transformations).
Superconformal transformations in SLFT are generated by the super energy-
momentum tensor $\hat{T}=S+2\theta T$ ($T$ is the ordinary energy-momentum
tensor and $S$ is a spin $3/2$ conserved current)
\begin{equation}
\label{A3}
\hat{T}=-\frac{1}{2 }D\Phi_{SL}\partial\Phi_{SL} +\frac{Q}{2}D
\partial \Phi_{SL},
\end{equation}
where
$D=\partial/\partial\theta+\theta \partial/\partial z$
is the covariant derivative.
As in every superconformal field theory \cite{P21} there are two
kinds of primary fields in SLFT: Neveu-Schwarz superfields
\begin{equation}
\label{A4}
\Phi _\alpha (Z,\bar{Z})=\phi_\alpha (z,\bar{z})+\theta
\psi_\alpha (z,\bar{z})+
\bar{\theta}\bar {\psi}_\alpha (z,\bar{z})-\theta\bar{\theta}
\tilde{\phi}_\alpha (z,\bar{z})\sim e^{\alpha \Phi_{SL} }
\end{equation}
 with
dimensions
\begin{equation}
\label{A5}
\Delta _\alpha =\frac{1}{2}\alpha (Q-\alpha ),
\end{equation}
(this fields are local with respect to fermionic current $S(z)$)
and Ramond fields 
\begin{equation}
\label{A6}
R_\alpha^{\epsilon}
=\sigma^{(\epsilon)}\phi_\alpha\sim \sigma ^{(\epsilon
)}e^{\alpha \phi_{SL}},
\end{equation}
where $\sigma ^{(\epsilon)} $ are so called twist fields
(they are quite similar to the
spin and disorder fields in 2d Izing model) with dimension $1/16$, so that 
the total dimension of $R_\alpha^{\epsilon}$ is
\begin{equation}
\label{A7}
\Delta _{[\alpha]}=\frac{1}{16}+\frac{1}{2}\alpha(Q-\alpha).
\end{equation}
The characteristic feature of Ramond fields is their nontrivial
($Z_2$)
monodromy with respect to fermionic current $S(z)$. All the other fields
of the theory can be obtained from Neveu-Schwarz (Ramond) primary fields
via the action of Neveu-Schwarz (Ramond)
algebra generators $L_n,S_m $, $n\in Z, m\in Z+1/2$ ($n\in Z, m \in Z$),
which are the Lorants coefficients of the energy-momentum tensor $T(z)$
and fermionic current $S(z)$ respectively. The commutation relations of the
Neveu-Schwarz-Ramond algebra have the form
\begin{equation}
\label{A8}
\left\{S_k, S_l\right\}=2
L_{k+l}+\frac{\hat{c}}{2}(k^2-1/4)\delta _{k+l,0},
\end{equation}
\begin{equation}
\label{A9}
\left[L_n, S_k\right]=\frac{1}{2}(n-2k)S_{n+k},
\end{equation}
\begin{equation}
\label{A10}
[L_n, L_m]=(n-m)L_{n+m}+\frac{\hat{c}}{8}(n^3-n)\delta _{n+m,0}.
\end{equation}
In our case of SLFT the central charge $\hat{c}$ of NSR algebra is equal to
\begin{equation}
\label{A11}
\hat{c}=1+2Q^2.
\end{equation}

Zero modes of fermionic currents $S_0$ and $\bar{S_0}$ act on Ramond
fields $R_\alpha ^{\epsilon }$
as follows:
\begin{equation}
\label{A12}
S_0 R_\alpha ^\epsilon=i\beta e^{-i\epsilon \pi/4}R_\alpha
^{-\epsilon },
\end{equation}
where 
\begin{equation}
\label{A13}
\beta = \frac{1}{\sqrt{2}}(\frac{Q}{2}-\alpha ).
\end{equation}

Further discussion is closely parallel to that of \cite{P17}.
Due to Liouville reflection the fields $\Phi_{\alpha}$ and
$\Phi_{Q-\alpha}$ are not independent (the same is
true for $R_{\alpha}^{\epsilon}$ and
 $R_{Q-\alpha}^{\epsilon }$) so that we can restrict
 the variation range of parameter $\alpha$ to be
 $\alpha\leq Q/2$.
It follows from superconformal symmetry, that the two-point
functions have the form:
\begin{equation}
\label{A14}
\langle\Phi _{\alpha _1} (Z_1,\bar{Z}_1)\Phi _{\alpha _2}
(Z_2,\bar{Z}_2)\rangle=\Delta (\alpha _1-\alpha
_2)(Z_{12} \bar{Z}_{12})^{-2\Delta _{\alpha _1}},
\end{equation}
\begin{equation}
\label{A15}
\langle R _{\alpha _1}^{\epsilon _1} (z_1,\bar{z}_1)R _{\alpha _2}
^{\epsilon _2}(z_2,\bar{z}_2)\rangle=\delta ^{\epsilon _1,\epsilon
_2}\Delta (\alpha _1-\alpha
_2)(z_{12}\bar{z}_{12})^{-2\Delta _{[\alpha _1]}},
\end{equation}
 where $\Delta(\alpha_1-\alpha_2)$ is some generalized function,
 which, according to the superconformal symmetry is nonzero only
 if $\alpha_1-\alpha_2=0$, and  the superdistance
 $Z_{12}=z_{12}-\theta_{1}
\theta_{2} $. Only at the end of the paper we'll restore
the proportionality coefficients in (\ref{A4}), (\ref{A6}) and extend the
variation range of parameter
$\alpha$ in order to obtain so called "reflection amplitudes".
As usual, the form of three point functions are restricted up to
some numerical coefficients by the superconformal invariance:
\begin{eqnarray}
\label{A16}
&& \qquad\langle\Phi _{\alpha _1} (Z_1,\bar{Z}_1)\Phi _{\alpha _2}
(Z_2,\bar{Z}_2) \Phi _{\alpha _3} (Z_3,\bar{Z}_3)\rangle =\nonumber \\
&& \quad =\left\{ C_{(\alpha_1),(\alpha _2),(\alpha _3)}+
\Theta\bar{\Theta }\tilde{C}_{(\alpha _1),(\alpha _2),(\alpha _3)}
\right\}(Z_{12}\bar{Z}_{12})^{\lambda  _3}
(Z_{13}\bar{Z}_{13})^{\lambda  _2}(Z_{23}\bar{Z}_{23})^{\lambda_1},
\end{eqnarray}
\begin{eqnarray}
\label{A17}
&&\langle \Phi _{\alpha _3}(Z_3,\bar{Z}_3) R _{\alpha _1}^{\epsilon _1}
(z_1,\bar{z}_1)R _{\alpha _2}^{\epsilon
_2}(z_2,\bar{z}_2)\rangle=(z_{12}\bar{z}_{12})^{\lambda _3}
(z_{13}\bar{z}_{13})^{\lambda_2}(z_{23}\bar{z}_{23})^{\lambda
_1}\times\nonumber \\
&&\times \left[\delta _{\epsilon _1,\epsilon _2}\left(C^{\epsilon_1 }_{[\alpha
_1],[\alpha _2],(\alpha _3)}+
\tilde{C}^{\epsilon_1 }_{[\alpha
_1],[\alpha _2],(\alpha _3)}
\tilde{C}^{\epsilon_1 }_{[\alpha
_1],[\alpha _2],(\alpha _3)}
\theta_3\bar{\theta}_3
|\frac{z_{31}z_{32}}{z_{12}}|\right)+\right.\nonumber\\
&&\left.+\delta _{\epsilon _1+\epsilon
_{2},0}\left(d^\epsilon_{[\alpha _1],[\alpha _2],(\alpha _3)}
 \left( \frac{z_{31}z_{32}}{z_{12}} \right)^{1/2}\theta_3+
 \bar{d}^\epsilon_{[\alpha _1],[\alpha _2],(\alpha _3)}
 \left( \frac{\bar{z}_{31}\bar{z}_{32}}{\bar{z}_{12}} \right)^
 {1/2}\bar{\theta}_3  \right)\right].
\end{eqnarray}
where $\lambda_{i}=2\Delta_{i}-\Delta_{1}-\Delta_{2}-\Delta_{3}$
, $i=1,2,3$ ($\Delta_{i}$ is the dimension of $i$-th field in the 
correlation function), $\Theta$ is the "superprojective
invariant" of three points
\begin{equation}
\label{A18}
\Theta=\left(z_{12}z_{13}z_{23} \right)^{-1/2}\left(z_{23}\theta _1
+z_{31}\theta _2+z_{12}\theta _3 -\frac{1}{2}
\theta _1\theta _2\theta _3 \right).
\end{equation}
Supersymmetry allows one to express coefficients 
$\tilde{C}^{\epsilon }_{[\alpha
_1],[\alpha _2],(\alpha _3)}$ and $d_{[\alpha
_1],[\alpha _2],(\alpha _3)}^\epsilon $  via
$C_{[\alpha_1],[\alpha _2],(\alpha _3)}^\epsilon $ as follows
\begin{equation}
\label{A19}
\tilde{C}_{[\alpha_1],[\alpha _2],(\alpha _3)}^\epsilon =
i\epsilon\left[\left( \beta_1^2+\beta_2^2\right) C_{[\alpha_1],[\alpha _2],
(\alpha _3)}^\epsilon -2\beta_1\beta_2C_{[\alpha_1],[\alpha _2],(\alpha _3)}
^{-\epsilon}\right],
\end{equation}
\begin{equation}
\label{A20}
d_{[\alpha_1],[\alpha _2],(\alpha _3)}^\epsilon=ie^{-\frac{i\pi\epsilon}{4}}
\left[ \beta_2C_{[\alpha_1],[\alpha _2],(\alpha _3)}^\epsilon -
\beta_1C_{[\alpha_1],[\alpha _2],(\alpha _3)}^{-\epsilon}
\right].
\end{equation}
The numerical coefficients $C,\tilde{C},d,\tilde{d}$ are called
structure constants and their calculation is the main purpose
of this work.

\section{\bf Degenerated Fields and Four Point Correlation
Functions}
\vspace{3mm}

For some special values of parameter $\alpha$  the primary fields
$\phi_{\alpha}$ ($R^\epsilon_{\alpha}$) become degenerated. This means, that
corresponding module Verma, i.e. the space, obtained with the 
help of successive actions by operators $S_{k}$,$L_{n}$,
$\bar{S_{k}}$, $\bar{L_{n}}$ with $k,n<0$ on primary fields
$\phi_{\alpha}$, ($R_{\alpha}^{\epsilon}$), contains
"null vector", i.e. some field $\chi_{\Delta+L}$ with properties
\begin{equation}
\label{A21}
L_{n}\chi_{\Delta+L}=S_{k}\chi_{\Delta+L}=0,\,\, {\rm for}\,\,
n,k>0,\,
L_{0}\chi_{\Delta+L}=(\Delta+L)\chi_{\Delta+L} ,
\end{equation}
where $\Delta$ is the conformal dimension of the field $\phi_{\alpha}$
($R_{\alpha}^{\epsilon}$) and the level of degeneracy $L$ is some
integer or half integer. To obtain an irreducible module, one
has to factorize the Verma modul over all submodules, generated
by the null vectors. In the field theory language this means, that 
we must put
\begin{equation}
\label{A22}
\chi_{\Delta+L}=0.
\end{equation}
Let us consider an example of degeneration, which plays an important
role in further discussion. The Ramond field 
$R_{\alpha}^{\epsilon}$ is degenerated
at the level $L=1$, if $\alpha=-b/2$, or $\alpha=-1/2b$.
The corresponding null vector has the form
\begin{equation}
\label{A23}
\chi=(\kappa L_{-1}-S_{-1}S_{0})R_{\alpha}^{\epsilon}=0,
\end{equation}
where
\begin{equation}
\label{A24}
\kappa =\left\{
\begin{array}{ll}
\displaystyle 1+\frac{1}{2b^2},& {\rm if}\quad\displaystyle  \alpha
=-\frac{b}{2},\\
&\\
\displaystyle 1+\frac{2b^2}{2},& {\rm if}\quad \displaystyle \alpha
=-\frac{1}{2b}.
\end{array}\right.
\end{equation}

Below we'll mainly consider the case $\alpha=-b/2$. To find
corresponding formulae for the case $\alpha=-1/2b$ one simply
has to replace $b\leftrightarrow 1/b$.
Following to \cite{P17}, it is easy to show, that the correlation function
$\langle \Phi_{\alpha _{3}}\Phi_{\alpha_2}
R_{\alpha}^{\epsilon}R_{\alpha_1}^{\epsilon_1}\rangle$, as a consequence
of eq. (\ref{A22}), satisfies the  following differential equation:
\begin{eqnarray}
\label{A25}
&&\kappa \frac{\partial}{\partial x_1}\langle \Phi
_{\alpha_ 3}\left(Z_3,\bar{Z}_3\right)\Phi_{\alpha _2}\left(Z_2,\bar{Z}_2\right)
R^{\epsilon }_{\alpha }\left(z,\bar{z}\right)
R^{\epsilon _1}_{\alpha_1}\left(z_1,\bar{z}_1\right) \rangle=\nonumber \\
&&= \sum_{i=2}^{3}\sqrt{\frac{(z_i-z)(z_i-z_1)}{z-z_1}}
\left\{-\frac{2\Delta _{\alpha _i}\theta _i}{(z-z_i)^2}
+\frac{1}{z-z_i}\hat{Q}_i\right\}\times\nonumber \\
&&\times \langle \Phi_{\alpha_ 3}\left(Z_3,\bar{Z}_3\right)\Phi_{\alpha
_2}\left(Z_2,\bar{Z}_2\right)S_0
R^{\epsilon }_\alpha \left(z,\bar{z}\right)
R^{\epsilon _1}_{\alpha_1}\left(z_1,\bar{z}_1\right) \rangle+\nonumber \\
&&+\frac{\beta ^2}{2(z-z_1)}\langle \Phi
_{\alpha_ 3}\left(Z_3,\bar{Z}_3\right)\Phi_{\alpha _2}
\left(Z_2,\bar{Z}_2\right)
R^{\epsilon }_\alpha \left(z,\bar{z}\right)
R^{\epsilon _1}_{\alpha_1}\left(z_1,\bar{z}_1\right) \rangle- \nonumber \\
&&-\frac{i\epsilon }{z-z_1}\langle \Phi
_{\alpha_ 3}\left(Z_3,\bar{Z}_3\right)\Phi_{\alpha _2}
\left(Z_2,\bar{Z}_2\right)S_0
R^{\epsilon }_\alpha \left(z,\bar{z}\right)S_0
R^{\epsilon _1}_{\alpha_1}\left(z_1,\bar{z}_1\right) \rangle,
\end{eqnarray}
where $\hat{Q}_i=\partial/\partial\theta_i-\theta_i
\partial/\partial z_i$, two
 possible choices of $\alpha$ and corresponding $\kappa$
are given by the eq.  (\ref{A24}) and $\beta$ can be
expressed via $\alpha$ with the help of eq.(\ref{A13}).
Equation (\ref{A25}) (and the analogous differential equation over complex
conjugate variables) makes it possible to obtain correlation
function
$\langle \Phi_{\alpha_ 3}\Phi_{\alpha _2}R^{\epsilon }_\alpha
R^{\epsilon _1}_{\alpha_1} \rangle$.
In fact,
due to superconformal symmetry it's quite sufficient to obtain
correlation functions
\begin{equation}
\label{A26}
G^{\epsilon }(z,\bar{z})=\lim_{|z_3 \bar{z_3}| \rightarrow \infty}
|z_3\bar{z}_3|^{2\Delta_{\alpha_3}}\langle \phi
_{\alpha_ 3}\left(z_3,\bar{z}_3\right)
\phi_{\alpha _2}\left(1\right)
R^{\epsilon }_\alpha \left(z,\bar{z}\right)
R^{\epsilon }_{\alpha_1 }\left(0\right) \rangle
\end{equation}
and
\begin{equation}
\label{A27}
H^{\epsilon }(z,\bar{z})=\lim_{|z_3 \bar{z_3}|\rightarrow \infty}
|z_3\bar{z}_3|^{2\Delta_{\alpha_3}}
\langle \phi
_{\alpha_ 3}\left(z_3,\bar{z}_3\right)
\psi_{\alpha _2}\left(1\right)
S_0 R^{\epsilon }_\alpha \left(z,\bar{z}\right)
R^{\epsilon }_{\alpha_1 }\left(0\right) \rangle ,
\end{equation}
Rewriting (\ref{A25}) in component language and adjusting coordinates
appropriately we obtain
\begin{equation}
\label{28}
\left(\kappa \frac{\partial }{\partial z}-\frac{\beta ^2}{2 z }\right)
G^{\epsilon }\left(z,\bar{z} \right)=\frac{\beta \beta_1
}{z}
\quad G^{-\epsilon}\left( z,\bar{z}\right)-\frac{ 1
}{\sqrt{z(1-z)}}H^{\epsilon }(z,\bar{z}),
\end{equation}

\begin{eqnarray}
\label{29}
&&\left(\kappa \frac{\partial }{\partial z}-\frac{\beta ^2}{2 z }\right)
H^{\epsilon }\left(z,\bar{z} \right)=-\frac{\beta \beta_1
}{z }
\quad H^{-\epsilon }(z,\bar{z})+\frac{2\Delta_{\alpha_2}  \beta^2
}{(1-z) \sqrt{z(1-z)}}
G^{\epsilon }\left(z,\bar{z}\right)-\nonumber \\
&&-\frac{ \beta^2
}{ \sqrt{z(1-z)}}\left(\gamma -z
\frac{\partial }{\partial z}\right)G^{\epsilon}(z,\bar{z}),
\end{eqnarray}
where $\gamma
=\Delta_{\alpha_3}-\Delta_{\alpha_2}-\Delta_{[\alpha]}-\Delta_{[\alpha_1]}$,
$\beta_1=(Q/2-\alpha_1)/\sqrt{2}$.
It's
convenient to introduce new functions
$G_{\epsilon}(z,\bar{z})$ and $H_{\epsilon }(z,\bar{z})$ as
follows:
\begin{equation}
\label{A30}
G^\epsilon (z,\bar{z})=G_+(z,\bar{z})+\epsilon G_-(z,\bar{z}),
\end{equation}
\begin{equation}
\label{A31}
H^\epsilon (z,\bar{z})=H_+(z,\bar{z})+\epsilon H_-(z,\bar{z}),
\end{equation}
The functions $G_{\epsilon}
(z,\bar{z})$ and $H_{\epsilon }(z,\bar{z})$ obey the following
system of differential equations:
\begin{equation}
\label{A32}
\left(\kappa \frac{\partial }{\partial z}-\frac{\beta ^2}{2z }\right)
G_{\epsilon }\left(z,\bar{z} \right)=\frac{\beta \beta_1\epsilon
}{z}
\quad G_{\epsilon}\left( z,\bar{z}\right)-\frac{ 1
}{\sqrt{z(1-z)}}H_{\epsilon }(z,\bar{z}),
\end{equation}

\begin{eqnarray}
\label{A33}
&&\left(\kappa \frac{\partial }{\partial z}-\frac{\beta ^2}{2z}\right)
H_{\epsilon }\left(z,\bar{z} \right)=\\
 &&\qquad = -\frac{\beta \beta_1\epsilon}{z}
H_{\epsilon }(z,\bar{z})
-\frac{\beta^2
}{\sqrt{z(1-z)}}\left(\gamma -\frac{2\Delta
_{\alpha_2}}{1-z}-z\frac{\partial}{\partial z} \right)
G_{\epsilon }(z,\bar{z}).\nonumber
\end{eqnarray}

Now, it's not difficult to exclude $H_{\epsilon }(z,\bar{z})$
and obtain the following second order linear differential equation
(in fact it coincides with Gauss hypergeometric equation):
\begin{equation}
\label{A34}
z(1-z)\frac{\partial^2}{\partial z^2}U_\epsilon
(z,\bar{z})-\left[ c_\epsilon -(a_\epsilon +b_\epsilon
+1)z\right]\frac{\partial}{\partial z}U_\epsilon
(z,\bar{z})-a_\epsilon b_\epsilon U_\epsilon(z,\bar{z})=0,
\end{equation}
where
\begin{equation}
\label{A35}
G_\epsilon (z,\bar{z})=(z \bar{z})^{\alpha _\epsilon
}\left[ (1-z)(1-\bar{z})\right]^{\beta _\epsilon }U_\epsilon
(z,\bar{z}),
\end{equation}
\begin{eqnarray}
\label{A36}
&&\alpha _\epsilon =\frac{1}{4}\left(\frac{1}{2}+b^2+b\epsilon
(Q-2\alpha_1 ) \right);\quad \beta _\epsilon
=\frac{1}{4}\left(1+b^2+b\epsilon (Q-2\alpha
_2)\right);\\
&&a_\epsilon =\frac{1}{4}\left(1+b\epsilon (3Q-2\alpha
_1-2\alpha _2-2\alpha _3) \right);\nonumber \\
&&b_\epsilon =\frac{1}{4}\left(1+b\epsilon (Q-2\alpha
_1-2\alpha _2+2\alpha _3) \right);\nonumber \\
&&c_\epsilon =\frac{1}{2}\left(1+b\epsilon (Q-2\alpha
_1) \right).\nonumber
\end{eqnarray}
Taking into account the mutual locality of the fields
$R^\epsilon_\alpha$, $R_{\alpha_{1}}^\epsilon,$
$\phi_{\alpha_{1}}$, $\phi_{\alpha_{2}}$
and that $U_{\epsilon }(z,\bar{z})$ obeys the same differential
equation also over the variable $\bar{z}$ it is straightforward to
obtain following expression:
\begin{eqnarray}
\label{A37}
&&G_\epsilon(z,\bar{z})=\left(z\bar{z}\right)^{\alpha
_\epsilon}\left[(1-z)(1-\bar{z})\right]^
{\beta _\epsilon}\times\nonumber \\
&&\times \left\{ g_\epsilon |F(a_\epsilon ,b_\epsilon
,c_\epsilon ,z)|^2+
\tilde{g}_\epsilon (z\bar{z})^{1-c_\epsilon }
|F(1+a_\epsilon - c_\epsilon, 1+b_\epsilon -c_\epsilon,
2-c_\epsilon, z)|^2\right\},
\end{eqnarray}
where
\begin{equation}
\label{A38}
\tilde{g}_\epsilon =-g_\epsilon \frac{\Gamma^2 (c_\epsilon
)\gamma (1-a_\epsilon )\gamma (1-b_\epsilon )}{\Gamma^
2(2-c_\epsilon)\gamma (c_\epsilon -a_\epsilon )\gamma (c_\epsilon
-b_\epsilon)},
\end{equation}
and $g_{\epsilon }$ are some constants to be defined later.
For our purposes it is useful also the following equivalent
expression for $G_{\epsilon}(z,\bar{z})$, which makes
$z\rightarrow 1$ asymptotics of the correlation function $G_{\epsilon}
(z,\bar{z})$ transparent:
\begin{eqnarray}
\label{A39}
&& G_\epsilon(z,\bar{z})=\left(z\bar{z}\right)^{\alpha _\epsilon
}\left[(1-z)(1-\bar{z})\right]^{\beta _\epsilon
}\left\{ f_\epsilon |F(a_\epsilon ,b_\epsilon
,a_\epsilon +b_\epsilon +1-c_\epsilon ,1-z)|^2+\right.\nonumber \\
&&\left.+\tilde{f}_\epsilon
\left((1-z)(1-\bar{z})\right)^{c_\epsilon -a_\epsilon -b_\epsilon }
|F(c_\epsilon -a_\epsilon,
c_\epsilon-b_\epsilon,c_\epsilon+1-a_\epsilon -b_\epsilon,
1- z)|^2\right\},
\end{eqnarray}
where
\begin{equation}
\label{A40}
f_\epsilon =g_\epsilon \frac{\gamma (c_\epsilon)
\gamma(c_\epsilon -a_\epsilon -b_\epsilon )}{\gamma (c_\epsilon
-a_\epsilon )\gamma (c_\epsilon -b_\epsilon )};\quad
\tilde{f}_\epsilon =g_\epsilon \frac{\gamma (c_\epsilon)
\gamma(a_\epsilon +b_\epsilon -c_\epsilon )}{\gamma (a_\epsilon
)\gamma (b_\epsilon )}.
\end{equation}

\section{\bf Three Point Functions and Reflection Amplitudes}
\vspace{3mm}

Expressions (\ref{A37}) and (\ref{A39}) show, that operator product
expansion of
the field $R_{-b/2}^{\epsilon }$ with arbitrary other primary
field (from the NS or R-sector) contains only finite number
of primary fields (see eq.(\ref{A77})). In fact, this property holds for all
degenerated primary fields, which are characterized by the
following spectrum of parameter $\alpha $:
\begin{equation}
\label{A41}
\alpha _{n,m}=\frac{1-n}{2}b+\frac{1-m}{2b},
\end{equation}
where $n$, $m$ are positive integers and $n-m=0(mod2)$
($n-m=1(mod2)$), if the corresponding field is from NS-sector
(R-sector).
The above mentioned property makes it possible (and convenient) to
choose usual in CFT discrete unit normalization for degenerated
fields instead of continuous normalization (\ref{A14}), (\ref{A15}) .
Investigating $z\rightarrow 0$ and $z\rightarrow1$ singularities
of correlation functions (\ref{A37}), (\ref{A39}) and identifying
intermediate states (see \cite{P17}) we see that the following operator
product expansions are valid:
\begin{eqnarray}
\label{A77}
R_{-b/2}^{\epsilon}(z,\bar{z})
R_{\alpha_1}^{\epsilon}(0)&=&
\sum_{\sigma=\pm 1}(z\bar{z})^{\Delta_{\alpha_1+\sigma b/2}-
\Delta_{[-b/2]}-\Delta_{[\alpha_1]}}
\left(
C_{[-b/2],[\alpha_1]}^{\epsilon,(\alpha_1+\sigma b/2)}
\phi_{\alpha_1+\sigma b/2}(0)+\right.\nonumber\\
&+&\left. |z|
(2\Delta_{\alpha_1+\sigma b/2})^{-2}
\tilde{C}_{[-b/2],[\alpha_1]}^{\epsilon,(\alpha_1+\sigma b/2)}
\tilde{\phi}_{\alpha_1+\sigma b/2}(0)+\ldots
\right)\\
R_{-b/2}^{\epsilon}(z,\bar{z})\phi_{\alpha_2}(0)&=&
\sum_{\sigma=\pm 1}(z\bar{z})^{\Delta_{[\alpha_2+\sigma b/2]}-
\Delta_{[-b/2]}-\Delta_{\alpha_2}}
\left(
C_{[-b/2],(\alpha_2)}^{\epsilon,([\alpha_2+\sigma b/2]}
R^{\epsilon}_{[\alpha_2+\sigma b/2]}(0)+\ldots\right).\nonumber
\end{eqnarray}
Let me note that owing to self-consistent normalization in NS- and 
R-sectors (\ref{A14},\ref{A15})
 $C_{[-b/2],[\alpha]}^{\epsilon,(\alpha_1\pm b/2)}=
C_{[-b/2],(\alpha\pm b/2)}^{\epsilon,[\alpha]}$.
Taking into account (\ref{A77}) it is easy to connect the constants 
$
g_\epsilon,\tilde{g}_{\epsilon},
f_\epsilon,\tilde{f}_{\epsilon}
$
from (\ref{A37},\ref{A39}) with the structure constants:
\begin{eqnarray}
\label{A42}
&&g_+=C^{\epsilon,(\alpha_1+b/2) }_{[-b/2],[\alpha _1]}
C_{(\alpha _1+b/2),(\alpha _2),(\alpha _3)};\nonumber \\
&& -\left( 2\Delta _{\alpha_1-b/2}\right)^2\tilde{g}_+
=\tilde{C}^{\epsilon,(\alpha_1-b/2)} _{[-b/2],[\alpha_1]}
\tilde{C}_{(\alpha _1-b/2),(\alpha _2),(\alpha _3)};\nonumber \\
&& \epsilon g_-=C^{\epsilon,(\alpha_1-b/2)}_{[-b/2],[\alpha _1]}
C_{(\alpha _1-b/2),(\alpha _2),(\alpha _3)};\nonumber \\
&& -\epsilon \left( 2\Delta _{\alpha
_1+b/2}\right)^2\tilde{g}_-=\tilde{C}^{\epsilon,(\alpha_1+b/2)}
_{[-b/2],[\alpha _1]}
\tilde{C}_{(\alpha _1+b/2),(\alpha _2),(\alpha _3)};\nonumber \\
&& f_++\epsilon
\tilde{f}_-=C^{\epsilon,[\alpha_2+b/2]}_{[-b/2],(\alpha_2)}
C^\epsilon _{[\alpha _1],[\alpha _2+b/2],(\alpha _3)};\nonumber \\
&&\tilde{f}_++\epsilon f_-=C^{\epsilon,[\alpha_2-b/2]}_{[-b/2],(\alpha_2)}
C^\epsilon _{[\alpha _1],[\alpha _2-b/2],(\alpha _3)}.
\end{eqnarray}
As the quantities $g_{\epsilon}$, $\tilde{g_{\epsilon}}$,
$f_{\epsilon}, \tilde{f_{\epsilon}}$ are connected via
eqs.(\ref{A38}),
(\ref{A40}), the relations (\ref{A42}) are, in fact, highly non
trivial functional relations for structure constants.
Before solving this functional relations, let us consider a special
case with $\alpha_{1}=-b/2$. As the field $\phi_{0}$, appearing
in s-channel, is the unit operator, in this case we have to put 
$g_{+}=\Delta (\alpha _2-\alpha _3)$, $g_{-}=f_{-}=\tilde{f_{-}}=0$
in (\ref{A42}). In this case it follows from
(\ref{A42}), that
\begin{eqnarray}
\label{A43}
&&C^{\epsilon,[\alpha +b/2]} _{[-b/2],(\alpha )}=
\epsilon \left[ \frac{\gamma (Qb)\gamma (\frac{(2\alpha -Q
)b}{2})}{\gamma(\frac{Qb}{2})\gamma (\alpha b)}\right]^{1/2},\nonumber \\
&&\nonumber \\
&&C^{\epsilon,[\alpha -b/2]}_{[-b/2],(\alpha )}=
\left[ \frac{\gamma (Qb)\gamma (\frac{(Q-2\alpha
)b}{2})}{\gamma(\frac{Qb}{2})\gamma ((Q-\alpha )b)}\right]^{1/2}.
\end{eqnarray}

Using (\ref{A19}), we easily obtain also the following structure constants:
\begin{eqnarray}
\label{A44}
&&\tilde{C}^{\epsilon,(\alpha -b/2)}_{[-b/2],[\alpha
]}=\frac{i}{2}(Q-\alpha +\frac{b}{2})^2
\left[ \frac{\gamma (Qb)\gamma (\alpha b+\frac{1}{2} -Q
b)}{\gamma\left(\frac{Qb}{2}\right)\gamma \left(\alpha
b+\frac{1}{2}-\frac{Qb}{2}\right)}\right]^{1/2},\nonumber \\
&&\tilde{C}^{\epsilon,(\alpha +b/2)}_{[-b/2],[\alpha
]}=\frac{i\epsilon}{2}(\alpha +\frac{b}{2})^2
\left[ \frac{\gamma (Qb)\gamma (\alpha b+\frac{1}{2} -\frac{Qb}{2}
}{\gamma\left(\frac{Qb}{2}\right)\gamma \left(\alpha
b+\frac{1}{2}\right)}\right]^{1/2}.
\end{eqnarray}

Inserting (\ref{A43}), (\ref{A44}) into (\ref{A42}), excluding
$g_{+}$, $\tilde{g_{+}}$
$g_{-}$, $\tilde{g_{-}}$ and $\tilde{C}_{(\alpha _1\pm
b/2),(\alpha _2),(\alpha _3)}$, we obtain the following functional
relation for $C_{(\alpha _1),(\alpha _2),(\alpha _3)}$:
\begin{eqnarray}
\label{A45}
&&\frac{C_{(\alpha _1+2b),(\alpha _2),(\alpha _3)}}
{C_{(\alpha _1),(\alpha _2),(\alpha _3)}}=
b^{4}
\left(
\frac{\gamma
        \left(\alpha_1b-\frac{Qb}{2}+2b^2
        \right)
        \gamma
        \left(
        \frac{2\alpha_1-Q}{2b}+3
        \right)}
        {\gamma
        \left(\alpha_1b-\frac{Qb}{2}
        \right)
        \gamma
        \left(\frac{2\alpha_1-Q}{2b}+1
        \right)}
\right)^{1/2}
       \times  \nonumber \\
&&\times \gamma \left( \alpha _1 b-\frac{Qb}{2}\right)\gamma
\left( \alpha _1 b+\frac{Qb}{2}-1\right)\gamma \left(\alpha
_1 b-Qb+1 \right)\gamma \left(\alpha _1 b\right)\times \\
&&\times\frac{\gamma \left(\frac{\delta _1b}{2}-Qb+1\right)\gamma
\left(\frac{\delta _1-Q}{2}b+1 \right)}{\gamma \left(\frac{\delta
b}{2} \right)\gamma \left(\frac{\delta-Q}{2}b \right)
\gamma \left( \frac{\delta _2+Q}{2}b\right)\gamma
\left(\frac{\delta _2}{2}b \right)\gamma
\left( \frac{\delta _3+Q}{2}b\right)\gamma
\left(\frac{\delta _3}{2}b \right)},\nonumber
\end{eqnarray}
where $\delta =\alpha _1+\alpha _2+\alpha _3$, $\delta _i=\delta
-2\alpha _i$.
An analogous functional relation, obtained via the substitution
$b\leftrightarrow 1/b$, is also valid
\begin{eqnarray}
\label{A46}
&&\frac{C_{(\alpha _1+2/b),(\alpha _2),(\alpha _3)}}
{C_{(\alpha _1),(\alpha _2),(\alpha _3)}}=
b^{-4}\left( \frac{\gamma \left(\frac{2\alpha
_1-Q}{2b}+\frac{2}{b^2}\right)\gamma \left(\alpha
_1b-\frac{Qb}{2}+3\right)}{\gamma \left(\frac{2\alpha
_1-Q}{2b}\right)\gamma \left( \alpha _1 b-\frac{Qb}{2}+1\right)}
\right)^{1/2}\times\nonumber \\
&&\times \gamma \left(\frac{2\alpha _1-Q}{2b}\right)\gamma
\left(\frac{2\alpha _1+Q}{2b}-1\right)\gamma \left(\frac{\alpha
_1-Q}{b}+1\right)\gamma \left(\frac{\alpha _1}{b}\right)\times \\
&&\times \frac{\gamma \left(\frac{\delta _1-2Q}{2b}+1\right)\gamma
\left(\frac{\delta _1-Q}{2b}+1 \right)}{\gamma \left(
\frac{\delta }{2b}\right)\gamma \left(\frac{\delta -Q}{2b}
\right)\gamma \left( \frac{\delta _2+Q}{2b}\right)\gamma
\left(\frac{\delta _2}{2b} \right)\gamma
\left( \frac{\delta _3+Q}{2b}\right)\gamma
\left(\frac{\delta _3}{2b} \right)},\nonumber
\end{eqnarray}

Functional relations (\ref{A45}), (\ref{A46}) (and the condition, that
$C_{(\alpha_1),(\alpha_2),(\alpha_3)}$ is a symmetric function
of $\alpha_1$, $\alpha_2$ and $\alpha_3$) determine the structure
constant $C_{(\alpha_1),(\alpha_2),(\alpha_3)}$ up to an overall
constant factor. The last statement follows from the fact, that
the ratio of any two solutions of (\ref{A45}),(\ref{A46}) would be
periodic with
two real (in general incommensurable, if $b^2$ is irrational)
periods $b$ and $1/b$. But such a function (under some natural 
assumptions of general character) must be constant.

The solution of (\ref{A45}), (\ref{A46}) can be expressed via the
function $\Upsilon
(x,Q)$, introduced by A.B. and Al.B.Zamolodchikovs in \cite{P16}
\begin{equation}
\label{A47}
\log\Upsilon(x,Q)=\int^{\infty}_{0}\frac{dt}{t}\left[
\left(\frac{Q}{2}-x \right)^2 e^{-t}-\frac{\sinh^2\left(
\frac{Q}{2}-x\right)\frac{t}{2}}{\sinh\frac{bt}{2}\sinh
\frac{t}{2b}}\right].
\end{equation}
Using the properties of $\Upsilon (x)$ (here and below we suppress
the parameter $Q$) \cite{P16}
\begin{equation}
\label{A48}
\Upsilon(x+b)=\gamma (bx)b^{1-2bx}\Upsilon(x),
\end{equation}
\begin{equation}
\label{A49}
\Upsilon(x+1/b)=\gamma (x/b)b^{2x/b-1}\Upsilon(x),
\end{equation}
one can easily check, that the following expression
is the solution of (\ref{A45}),(\ref{A46}):
\begin{eqnarray}
\label{A50}
&&C_{(\alpha_1),(\alpha _2),(\alpha _3)}=C_0\Upsilon _0
b^{(3Q-\alpha _1-\alpha _2-\alpha _3 )(1/b-b)}\times\nonumber \\
&&\times\prod_{i=1}^{3}\left[
\left(\gamma
\left(1+\frac{Q}{2b}-\frac{\alpha _i}{b} \right)\gamma
\left(\frac{Qb}{2}-\alpha _i b\right) \right)^{1/2} \prod_{\sigma \in{0,1}}
\frac{\Upsilon \left(\alpha _i+\frac{\sigma Q}{2}
\right)}{\Upsilon \left(\frac{\delta _i+\sigma Q}{2}\right)
\Upsilon \left(\frac{\delta -\sigma Q}{2}\right)}\right].
\end{eqnarray}
where $C_{0}$ is a constant, whose value depends on concrete choice
of normalizing function $\Delta(\alpha)$ in (\ref{A14}) and
\begin{equation}
\label{A51}
\Upsilon  _0=\frac{d\Upsilon (x)}{dx}|_{x=0}.
\end{equation}
Using (\ref{A42}) and (\ref{A50}), for the structure constant
$\tilde{C}_{(\alpha_1),(\alpha_2),
(\alpha_3)}$ we obtain
\begin{eqnarray}
\label{A52}
\displaystyle \tilde{C}_{(\alpha_1),(\alpha _2),(\alpha _3)}&=&
-2i C_0\Upsilon _0
b^{(3Q-\alpha _1-\alpha _2-\alpha _3 )(1/b-b)} \times\nonumber\\
&\times&
\displaystyle \prod_{i=1}^{3}\left[
\left(\gamma
\left(1+\frac{Q}{2b}-\frac{\alpha _i}{b} \right)\gamma
\left(\frac{Qb}{2}-\alpha _i b\right) \right)^{1/2}
\times\right. \\
&\times&
\left. \frac{\Upsilon \left(\alpha _i\right)
\Upsilon \left(\alpha _i+\frac{Q}{2} \right)}
{\Upsilon \left(\frac{\delta_i}{2}+\frac{1}{2b}\right)
\Upsilon \left(\frac{\delta_i}{2}+\frac{b}{2}\right)
\Upsilon \left(\frac{\delta}{2}-\frac{1}{2b}\right)
\Upsilon
\left(\frac{\delta}{2}-\frac{b}{2}\right)}\right]. \nonumber
\end{eqnarray}

Now, it is not difficult to find corresponding expressions for
the Ramond sector. Indeed, using (\ref{A40}), (\ref{A42}) and
(\ref{A43}) we get
\begin{equation}
\label{A53}
f_+=\left[\frac{\gamma \left(Qb\right)
\gamma \left( \frac{Q-2\alpha _1}
{2}b+\frac{1}{2}\right)}{\gamma \left(\frac{Qb}{2} \right)\gamma
\left(\alpha _1 b +\frac{1}{2}\right)}\right]^{1/2}
\frac{\gamma \left(\frac{2\alpha_2-Q}{2}b \right) C_{(\alpha_1+b/2),
(\alpha_2),(\alpha_3)}}{\gamma\left(\frac{2\delta_1-Q}{4}b+
\frac{1}{4}\right)
\gamma\left(\frac{Q-2\delta_2}{4}b+\frac{1}{4} \right)},
\end{equation}
\begin{equation}
\label{A54}
f_-=\left[\frac{\gamma \left(Qb\right)
\gamma \left((\alpha _1-Q)b
+\frac{1}{2}\right)}{\gamma \left(\frac{Qb}{2} \right)\gamma
\left( \frac{Q-2\alpha_1}{2}b +\frac{1}{2}\right)}\right]^{1/2}
\frac{\gamma \left(\frac{Q-2\alpha_2}{2}b \right) C_{(\alpha_1-b/2),
(\alpha_2),(\alpha_3)}}{\gamma\left(\frac{Q-2\delta_1}{4}b+
\frac{1}{4}\right)
\gamma\left(\frac{2\delta_2-Q}{4}b+\frac{1}{4} \right)}.
\end{equation}
Inserting (\ref{A53}), (\ref{A54}) into (\ref{A42}) and shifting the
parameter $\alpha_{2}\rightarrow \alpha_{2}-b/2$, we obtain:
\begin{eqnarray}
\label{A55}
&&\displaystyle C_{[\alpha_1],[\alpha _2],(\alpha _3)}^\epsilon
=\nonumber\\
&&= C_0\Upsilon _0
b^{(3Q-\delta )(1/b-b)-2}
\left(\frac{\gamma \left(\frac{Qb}{2}-\alpha _3 b\right)
\gamma \left(1+\frac{Q}{2b}-\frac{\alpha _3}{b}\right)}
{\gamma \left(\alpha _1 b-\frac{b^2}{2} \right)
\gamma \left(\alpha _2 b-\frac{b^2}{2} \right)
\gamma \left(\frac{\alpha _1}{b}-\frac{1}{2b^2} \right)
\gamma \left(\frac{\alpha _2}{b}-\frac{1}{2b^2} \right)
} \right)^{1/2} \times \nonumber \\
&&\displaystyle \times
\left[
\frac{\Upsilon \left(\alpha _1+\frac{b}{2}\right)
\Upsilon \left(\alpha _2+\frac{b}{2}\right)
\Upsilon \left(\alpha _1+\frac{1}{2b}\right)
\Upsilon \left(\alpha _2+\frac{1}{2b}\right)
\Upsilon \left(\alpha _3\right)
\Upsilon \left(\alpha _3+\frac{Q}{2}\right)
 }{\Upsilon \left(\frac{\delta _1}{2}\right)
 \Upsilon \left(\frac{\delta _1 +Q}{2}\right)
 \Upsilon \left(\frac{\delta _2}{2}\right)
 \Upsilon \left(\frac{\delta _2 +Q}{2}\right)
 \Upsilon \left(\frac{\delta _3 +b}{2}\right)
 \Upsilon \left(\frac{b\delta _3 +1}{2b}\right)
 \Upsilon \left(\frac{\delta -b}{2}\right)
 \Upsilon \left(\frac{b\delta -1}{2b}\right)}+\right. \nonumber \\
 &&\displaystyle \left.+\frac{\epsilon \Upsilon
 \left(\alpha _1+\frac{b}{2}\right)
\Upsilon \left(\alpha _2+\frac{b}{2}\right)
\Upsilon \left(\alpha _1+\frac{1}{2b}\right)
\Upsilon \left(\alpha _2+\frac{1}{2b}\right)
\Upsilon \left(\alpha _3\right)
\Upsilon \left(\alpha _3+\frac{Q}{2}\right)
 }{\Upsilon \left(\frac{\delta _1+b}{2}\right)
\Upsilon \left(\frac{\delta _2+b}{2}\right)
\Upsilon \left(\frac{b\delta _1 +1}{2b}\right)
\Upsilon \left(\frac{b\delta _2 +1}{2b}\right)
 \Upsilon \left(\frac{\delta _3}{2}\right)
 \Upsilon \left(\frac{\delta _3 +Q}{2}\right)
 \Upsilon \left(\frac{\delta}{2}\right)
 \Upsilon \left(\frac{\delta -Q}{2}\right)} \right].
\end{eqnarray}

Up to now we have assumed all $\alpha$'s to be restricted to
$\alpha \leq Q/2$, but the expressions (\ref{A50}), (\ref{A52}),
 (\ref{A55})
are defined out of
this region as well and it is interesting to note, that the
following relations are valid:
\begin{equation}
\label{A56}
C_{(\alpha _1),(\alpha _2),(\alpha _3)}=
C_{(Q-\alpha _1),(\alpha _2),(\alpha _3)},
\end{equation}
\begin{equation}
\label{A57}
\tilde{C}_{(\alpha _1),(\alpha _2),(\alpha _3)}=
\tilde{C}_{(Q-\alpha _1),(\alpha _2),(\alpha _3)},
\end{equation}
\begin{equation}
\label{A58}
C_{[\alpha _1],[\alpha _2],(\alpha _3)}^\epsilon =
\epsilon C_{[Q-\alpha _1],[\alpha _2],(\alpha _3)}^\epsilon =
C_{[\alpha _1],[\alpha _2],(Q-\alpha _3)}^\epsilon.
\end{equation}
Let us introduce normalizing coefficients in (\ref{A4}), (\ref{A6})
explicitly
\begin{equation}
\label{A59}
\exp\alpha \Phi _{SL}=\nu (\alpha )\Phi _{\alpha }.
\end{equation}
\begin{equation}
\label{A60}
\sigma ^{(\epsilon)} \exp\alpha \phi _{SL}=\rho (\alpha )R _{\alpha
}^\epsilon
\end{equation}
Similar to ordinary bosonic Liouville theory,
the correlation functions of exponential fields have poles if 
total charge $\sum \alpha_{i}=Q-nb$; $n\in 0,1,2,\ldots $.
The residues of these poles can be calculated via perturbative expansion
over cosmological constant $\mu $ (in fact only the term $\sim 
\mu^{n}$ contributes). In particular, if $n=0$ we have correlation
functions in a free theory ($\mu =0$) with background charge $Q$
at infinity. So, for the three point functions we have:
\begin{eqnarray}
\label{A61}
&&{\rm res}_{ \sum \alpha _i=Q} \langle e^{\alpha _1\Phi _{SL}}
e^{\alpha _2\Phi _{SL}} e^{\alpha _3\Phi _{SL}}\rangle =
\langle e^{\alpha _1\Phi _{SL}}
e^{\alpha _2\Phi _{SL}} e^{\left(Q-\alpha _1-\alpha _2 \right)
\Phi _{SL}}\rangle_{\mu =0} =\nonumber \\
&&=|Z_{12}|^{2\lambda _{3}}|Z_{13}|^{2\lambda
_{2}}|Z_{23}|^{2\lambda _{1}},
\end{eqnarray}
\begin{eqnarray}
\label{A62}
&&{\rm res}_{ \sum \alpha _i=Q} \langle e^{\alpha _1\phi
_{SL}}\sigma ^\epsilon
e^{\alpha _2\phi _{SL}}\sigma ^\epsilon  e^{\alpha _3\phi
_{SL}}\rangle =
\langle e^{\alpha _1\phi _{SL}}\sigma ^\epsilon
e^{\alpha _2\phi _{SL}}\sigma ^\epsilon  e^{\left(Q-\alpha _1
-\alpha _2 \right)\phi _{SL}}\rangle_{\mu =0} =\nonumber \\
&&=|z_{12}|^{2\lambda _{3}}|z_{13}|^{2\lambda
_{2}}|z_{23}|^{2\lambda _{1}},
\end{eqnarray}
The relations (\ref{A61}), (\ref{A62}) determine normalizing
functions $\nu (\alpha)$, $\rho (\alpha)$ up to constant parameters
$\kappa$, $C_{0}$:
\begin{equation}
\label{A63}
\nu (\alpha )=\left(2b^{2Q\left( 1/b-b\right)}C_0\right)^{-1/3}\kappa
^{\alpha-\frac{Q}{3}}\left( \frac{\gamma \left( \frac{\alpha
}{b}-\frac{Q}{2b}\right)}{\gamma \left( -\alpha b+\frac{Qb}{2}\right)}
\right)^{1/2},
\end{equation}

\begin{equation}
\label{A64}
\rho  (\alpha )=\left(2b^{2Q\left( 1/b-b\right)}C_0\right)^{-1/3}\kappa
^{\alpha-\frac{Q}{3}}\left( \frac{b^2 \gamma \left( \frac{\alpha
}{b}-\frac{1}{2b^2}\right)}{\gamma \left(1 -\alpha
b+\frac{b^2}{2}\right)}\right)^{1/2}.
\end{equation}
To express unknown constant $\kappa$, via cosmological
constant $\mu$ and coupling constant $b$, one can in a similar way
investigate the case $n=1$:
\begin{eqnarray}
\label{A65}
&&{\rm res}_{ \sum \alpha _i=1/b} \langle e^{\alpha _1\Phi _{SL}}
e^{\alpha _2\Phi _{SL}} e^{\alpha _3\Phi _{SL}}\rangle =\nonumber \\
&&=i\mu \int d^2 z_4 d^2\theta _4
\langle e^{\alpha _1\Phi _{SL}}
e^{\alpha _2\Phi _{SL}} e^{\left(\frac{1}{b}-\alpha _1-\alpha _2 \right)
\Phi _{SL}}e^{b\Phi _{SL}\left( Z_4,\bar{Z}_4\right)}\rangle_{\mu =0} =\nonumber \\
&&=-i\pi \mu \frac{\gamma \left( b\alpha _1+b\alpha
_2\right)}{\gamma \left(b\alpha _1 \right)\gamma \left(b\alpha _2\right)}
\Theta\bar{\Theta }|Z_{12}|^{2\lambda _{3}}|Z_{13}|^{2\lambda
_{2}}|Z_{23}|^{2\lambda _{1}}.
\end{eqnarray}
Using eq. (\ref{A65}) we easily obtain, that
 \begin{equation}
\label{A66}
\kappa =b^{b-1/b}\left(\frac{\pi \mu }{2}\gamma \left(
\frac{1+b^2}{2}\right) \right)^{-1/b}.
\end{equation}
From  (\ref{A50}), (\ref{A55}), (\ref{A63}), (\ref{A64}),
(\ref{A66}) we obtain final expressions for the structure
constants of exponential fields $\exp \alpha \phi_{SL}$,
$\sigma^{(\epsilon)}\exp \alpha \phi_{SL}$:
\begin{eqnarray}
\label{A67}
&& {\bf C}_{(\alpha_1),(\alpha _2),(\alpha _3)}=\frac{1}{2}\left(
\frac{\pi \mu }{2}\gamma
\left(\frac{1+b^2}{2}\right)b^{2-2b^2}\right)^{\frac{Q-\delta
}{b}}\times\nonumber \\
&&\times\Upsilon _0 \prod_{i=1}^{3}\prod_{\sigma \in{0,1}}
\frac{\Upsilon \left(\alpha _i+\frac{\sigma Q}{2}
\right)}{\Upsilon \left(\frac{\delta _i+\sigma Q}{2}\right)
\Upsilon \left(\frac{\delta -\sigma Q}{2}\right)},
\end{eqnarray}

\begin{eqnarray}
\label{A68}
&&\displaystyle \tilde{{\bf C}}_{(\alpha_1),(\alpha _2),(\alpha _3)}=
-i\left(\frac{\pi \mu }{2}\gamma
\left(\frac{1+b^2}{2}\right)b^{2-2b^2}\right)^{\frac{Q-\delta
}{b}}\Upsilon_0 \times \nonumber \\
&&\displaystyle \times
\prod_{i=1}^{3} \frac{\Upsilon \left(\alpha _i \right) \Upsilon
\left(\alpha _i+\frac{Q}{2} \right)} {\Upsilon
\left(\frac{\delta_i}{2}+\frac{1}{2b}\right) \Upsilon
\left(\frac{\delta_i}{2}+\frac{b}{2}\right) \Upsilon
\left(\frac{\delta}{2}-\frac{1}{2b}\right) \Upsilon
\left(\frac{\delta}{2}-\frac{b}{2}\right)},
\end{eqnarray}

\begin{eqnarray}
\label{A69}
&&\displaystyle {\bf C}_{[\alpha_1],[\alpha _2],(\alpha _3)}^\epsilon
=\frac{1}{2}\left(\frac{\pi \mu }{2}\gamma
\left(\frac{1+b^2}{2}\right)b^{2-2b^2}\right)^{\frac{Q-\delta
}{b}}\Upsilon_0\times \nonumber \\
&&\displaystyle \times
\left[
\frac{\Upsilon \left(\alpha _1+\frac{b}{2}\right)
\Upsilon \left(\alpha _2+\frac{b}{2}\right)
\Upsilon \left(\alpha _1+\frac{1}{2b}\right)
\Upsilon \left(\alpha _2+\frac{1}{2b}\right)
\Upsilon \left(\alpha _3\right)
\Upsilon \left(\alpha _3+\frac{Q}{2}\right)
 }{\Upsilon \left(\frac{\delta _1}{2}\right)
 \Upsilon \left(\frac{\delta _1 +Q}{2}\right)
 \Upsilon \left(\frac{\delta _2}{2}\right)
 \Upsilon \left(\frac{\delta _2 +Q}{2}\right)
 \Upsilon \left(\frac{\delta _3 +b}{2}\right)
 \Upsilon \left(\frac{b\delta _3 +1}{2b}\right)
 \Upsilon \left(\frac{\delta -b}{2}\right)
 \Upsilon \left(\frac{b\delta -1}{2b}\right)}+\right. \nonumber \\
 &&\displaystyle \left.+\frac{\epsilon \Upsilon
 \left(\alpha _1+\frac{b}{2}\right)
\Upsilon \left(\alpha _2+\frac{b}{2}\right)
\Upsilon \left(\alpha _1+\frac{1}{2b}\right)
\Upsilon \left(\alpha _2+\frac{1}{2b}\right)
\Upsilon \left(\alpha _3\right)
\Upsilon \left(\alpha _3+\frac{Q}{2}\right)
 }{\Upsilon \left(\frac{\delta _1+b}{2}\right)
\Upsilon \left(\frac{\delta _2+b}{2}\right)
\Upsilon \left(\frac{b\delta _1 +1}{2b}\right)
\Upsilon \left(\frac{b\delta _2 +1}{2b}\right)
 \Upsilon \left(\frac{\delta _3}{2}\right)
 \Upsilon \left(\frac{\delta _3 +Q}{2}\right)
 \Upsilon \left(\frac{\delta}{2}\right)
\Upsilon \left(\frac{\delta -Q}{2}\right)} \right].
\end{eqnarray}
Reflection properties (\ref{A56})-(\ref{A58}) now take the form:
\begin{equation}
\label{A70}
{\bf C}_{(\alpha _1),(\alpha _2),(\alpha _3)}=
G_{NS}(\alpha _1){\bf C}_{(Q-\alpha _1),(\alpha _2),(\alpha _3)},
\end{equation}
\begin{equation}
\label{A71}
\tilde{{\bf C}}_{(\alpha _1),(\alpha _2),(\alpha _3)}=
G_{NS}(\alpha _1){\bf C}_{(Q-\alpha _1),(\alpha _2),(\alpha _3)},
\end{equation}
\begin{equation}
\label{A72}
{\bf C}_{[\alpha _1],[\alpha _2],(\alpha _3)}^\epsilon =
G_{R}(\alpha _1)\epsilon {\bf C}^\epsilon _{[Q-\alpha _1],[\alpha _2],(\alpha _3)}=
G_{NS}(\alpha _3){\bf C}^\epsilon _{[\alpha _1],[\alpha _2],(Q-\alpha _3)},
\end{equation}
where
\begin{equation}
\label{A73}
G_{NS}(\alpha)=\frac{\nu (\alpha )}{\nu (Q-\alpha )}=\left(\frac{\pi \mu }
{2}\gamma \left(\frac{1+b^2}{2}\right)b^{1-b^2}\right)^{\frac{Q-2\alpha }
{b}}
\frac{b^2\gamma (\alpha  b-\frac{Qb}{2})}{\gamma \left(-\frac{\alpha }{b}
+\frac{Q}{2b} \right)},
\end{equation}
\begin{equation}
\label{A74}
G_R(\alpha)=\frac{\rho (\alpha )}{\rho (Q-\alpha )}=\left(\frac{\pi \mu }
{2}\gamma \left(\frac{1+b^2}{2}\right)b^{1-b^2}\right)^{\frac{Q-2\alpha }{b}}
\frac{\gamma \left(\frac{1}{2}-\frac{Qb}{2}+\alpha b\right)}
{\gamma \left(\frac{1}{2}
+\frac{Q}{2b}-\frac{\alpha }{b}\right)}.
\end{equation}
When $\alpha = Q/2+ip$; $p\in \cal R$ (only the states, corresponding
to such charges and their descendants contribute to one loop
partition
function \cite{P7}), the functions (\ref{A72}), (\ref{A73}) are called
Super-Liouville "reflection amplitudes"
\begin{equation}
\label{A75}
S_{NS}(P)=G_{NS}\left(\frac{Q}{2}+iP\right)= -
\left(\frac{\pi \mu }{2}\gamma \left(\frac{1+b^2}{2}\right)
b^{1-b^2}\right)^{-\frac{2iP}{b}}
\frac{\Gamma (1+iPb)\Gamma \left(1+\frac{iP}{b}\right)}{\Gamma (1-iPb)\Gamma
\left(1-\frac{iP}{b}\right)},
\end{equation}
\begin{equation}
\label{A76}
S_R(P)=G_R\left(\frac{Q}{2}+iP\right)=
\left(\frac{\pi \mu }{2}\gamma \left(\frac{1+b^2}{2}\right)
b^{1-b^2}\right)^{-\frac{2iP}{b}}
\frac{\Gamma \left(\frac{1}{2}+iPb\right)\Gamma \left(\frac{1}{2}+
\frac{iP}{b}\right)}{\Gamma \left(\frac{1}{2}-iPb\right)\Gamma
\left(\frac{1}{2}-\frac{iP}{b}\right)}.
\end{equation}
Reflection amplitudes $S_{NS}(p)$ and $S_{R}(p)$ have unit modules,
which means, that as in bosonic case we have complete reflection.

\vspace{1cm}

\noindent
{\large{\bf Acknowledgments}}

\noindent
I would like to thank A.Belavin for his stimulating
interest to this work and valuable discussions. I'm also
grateful to J.Ambjorn for interesting discussion and T.Hakobyan
for useful collaboration. Tis work was supported in part by
grant 211-5291 YPI of the German Bundesministerium fur
Forschung und Technologie and by grant INTAS-93-633.

\vspace{1cm}
\noindent
{\bf Note Added}

\noindent
After this work was completed I have learned that in
ref.\cite{P22} three point functions of $N=1$ SLFT were
proposed, extending the method of the paper \cite{P16} to the
supersymmetric case. It is worth noting, that the generalized
special functions $\Upsilon_1(x)$, $\Upsilon_2(x)$, introduced
by the authors of \cite{P22} and used in the expressions of the
three-point functions, in fact, can be expressed via the function
 $\Upsilon(x)$
as follows: $\Upsilon_1(2x)=\Upsilon(x)\Upsilon(x+Q/2)$,
$\Upsilon_2(2x)=\Upsilon(x+b/2)\Upsilon(x+1/2b)$. These relations
can be checked directly using the definitions. Performing
corresponding substitutions, it is easy to see, that the results,
of \cite{P22} coincide with the results, presented here in the
NS-sector, while in the case of R-sector there are some
differences.

\end{document}